\documentclass[runningheads]{llncs}
\usepackage[T1]{fontenc}

\usepackage{tabularx}
\usepackage{booktabs}
\usepackage{graphicx}
\usepackage{subfigure}
\usepackage{cite}
\usepackage{multirow}
\usepackage{xcolor}
\usepackage{amsmath,amsfonts,amssymb}

\usepackage{marvosym}
\usepackage{pifont} 
\usepackage{mathtools}
\usepackage{float}
\usepackage{amsmath}

\usepackage{color}

\usepackage{soul}

\usepackage{hyperref}
\let\svthefootnote\thefootnote
\newcommand\freefootnote[1]{%
  \let\thefootnote\relax%
  \footnotetext{#1}%
  \let\thefootnote\svthefootnote%
}

\usepackage{perpage} 
\MakePerPage{footnote} 

\begin{document}
%
\title{
Score-based Diffusion Model
 for Unpaired Virtual Histology Staining}
\titlerunning{ Diffusion for Unpaired Virtual Histology Staining}

\author{Anran Liu\inst{1}\textsuperscript{$*$}, Xiaofei Wang\inst{2}\textsuperscript{$*$},  Jing Cai \inst{1}\textsuperscript{\Letter}, Chao Li\inst{2,3,4}\textsuperscript{\Letter}}
\index{Liu Anran}
\index{Wang Xiaofei}
\index{Cai Jing}
\index{Li Chao}

\authorrunning{A. Liu et al.}

\institute{
Department of Health Technology and Informatics, Hong Kong Polytechnic University, China \and 
Department of Clinical Neurosciences, University of Cambridge, UK \and 
School of Science and Engineering, University of Dundee, UK \and
Department of Applied Mathematics and Theoretical Physics, University of
Cambridge, UK \\
\email{jing.cai@polyu.edu.hk}
\email{cl647@cam.ac.uk}
}
\maketitle              

\begin{abstract}
Hematoxylin and eosin (H\&E) staining visualizes histology but lacks specificity for diagnostic markers. Immunohistochemistry (IHC) staining provides protein-targeted staining but is restricted by tissue availability and antibody specificity. 
Virtual staining, i.e., computationally translating the H\&E image to its IHC counterpart while preserving the tissue structure, is promising for efficient IHC generation.
Existing virtual staining methods still face key challenges: 1) effective decomposition of staining style and tissue structure, 2) controllable staining process adaptable to diverse tissue and proteins, and 3) rigorous structural consistency modelling to handle the non-pixel-aligned nature of paired H\&E and IHC images.
This study proposes a mutual-information (MI)-guided score-based diffusion model for unpaired virtual staining. Specifically, we design 1) a global  MI-guided energy function that disentangles the tissue structure and staining characteristics across modalities, 2) a novel timestep-customized reverse diffusion process for precise control of the staining intensity and structural reconstruction, and 3) a local MI-driven contrastive learning strategy to ensure the cellular level structural consistency between H\&E-IHC images.
Extensive experiments demonstrate the our superiority over state-of-the-art approaches, highlighting its biomedical potential. 
Codes will be open-sourced upon acceptance.
\freefootnote{$*$ Equal contribution.}
\end{abstract}
\section{Introduction}
\label{sec:introduction}
Hematoxylin and eosin (H\&E) staining is the most common diagnostic practice to evaluate histology. However, H\&E staining cannot provide information on specific protein expressions \cite{fischer2008hematoxylin}, which can be targeted by immunohistochemistry (IHC) staining, offering sensitive diagnostic markers. However, IHC staining is limited by laborious and time-intensive procedures, especially when tissue samples are scarce. To mitigate these challenges, virtual staining has emerged as a promising approach to generate IHC images directly from H\&E images.

Traditional methods for virtual histology staining typically rely on pixel-wise color mapping to transfer histological features from one staining to another \cite{hanselmann2009toward,wieslander2021learning}. These methods often struggle to capture the complex relationships between different stainings. Advanced methods, particularly generative models such as generative adversarial networks (GANs), have emerged to more effectively generate IHC staining based on H\&E images. Generally, in developing generative models, major challenges persist due to the inevitable discrepancies in tissue and cellular morphology between H\&E and IHC images, as they are derived from different sections,  though usually from the same tissue \cite{li2023adaptive}.

To tackle this challenge, unpaired GAN-based methods \cite{li2024exploiting,li2023adaptive} were proposed. For instance, Qin \textit{et al.} \cite{chen2024pathological} attempted to incorporate molecular-level semantic information to mitigate the impact caused by spatial inconsistencies. Despite impressive performance \cite{liu2021unpaired}, GAN-based models often face issues such as mode collapse and instability during training \cite{zhang2018convergence}. Notably, the entanglement of shared information (i.e., structural features) and specific information (i.e., staining style) during the generation complicates the precise generation control, leading to staining bias or structural information loss.

Recently, diffusion models \cite{moghadam2023morphology} have advanced controllable image generation, achieving high stability, broad mode coverage, and superior sample fidelity in both natural and biomedical images\cite{wang2024cross,xiang2023ddm,he2024pst}. 
For example, Luo \textit{et al}. \cite{luo2024target} proposed to guide the unpaired MRI-CT generation through a classifier-based filter, aiming to recover specific details in the target domain.
Similarly,  Kataria \textit{et al}. \cite{kataria2024staindiffuser} developed StainDiffuser, a multi-task architecture using a dual diffusion model to transfer information through a shared encoder and convert H\&E to IHC images. However, existing methods assume strict pixel-wise alignment, contradicting the unpaired nature of real-world  H\&E and IHC images. Hence, it remains an unmet need to design unpaired diffusion models for effective virtual staining.   



Despite advances in natural images\cite{su2022dual,zhao2022egsde,xu2024cyclenet},  developing diffusion models for unpaired virtual histology staining presents unoque challenges, as it requires: 1) effectively disentangling specific staining information and shared structure information between H\&E and IHC images, to assign appropriate staining patterns to specific tissue regions, 2) a customizable staining process adjusting to various tissue and proteins, and 3) a rigorous structural consistency between IHC and H\&E images to reliably preserve tissue morphological details. A prior diffusion-based model, PST-Diff \cite{he2024pst}, trains two parallel diffusion models for unpaired H\&E and IHC images by imposing constraints within the noise latent space. 
However, this framework has high computational complexity with style and structure separately modelled, leading to risks of unintended mixing of structural and staining information, as consistency constraints are limited in the latent space.

To address the challenges, we propose a mutual information (MI)-guided unpaired virtual staining via the score-based diffusion model (MIU-Diff). To the best of our knowledge, this is the first study for unpaired virtual histology staining via energy-guided diffusion modelling. Our main contribution includes:
\begin{itemize}
\item We develop a novel score-based diffusion framework, with a global MI-guided energy function for dynamic disentangling the staining characteristics and structure details across modalities of H\&E and IHC images.

\item We propose to solve a novel timestep-adaptive reverse-time stochastic differential equation (SDE)
in the diffusion process to precisely control staining intensity and structural reconstruction.

\item We design a local MI-guided contrastive learning strategy to maintain cellular-level structural consistency between paired images.
\end{itemize}

Our extensive experiments on the two public datasets demonstrate that our method outperforms other state-of-the-art (SOTA) methods

\begin{figure}[t]
    \centering
    \includegraphics[width=\linewidth]{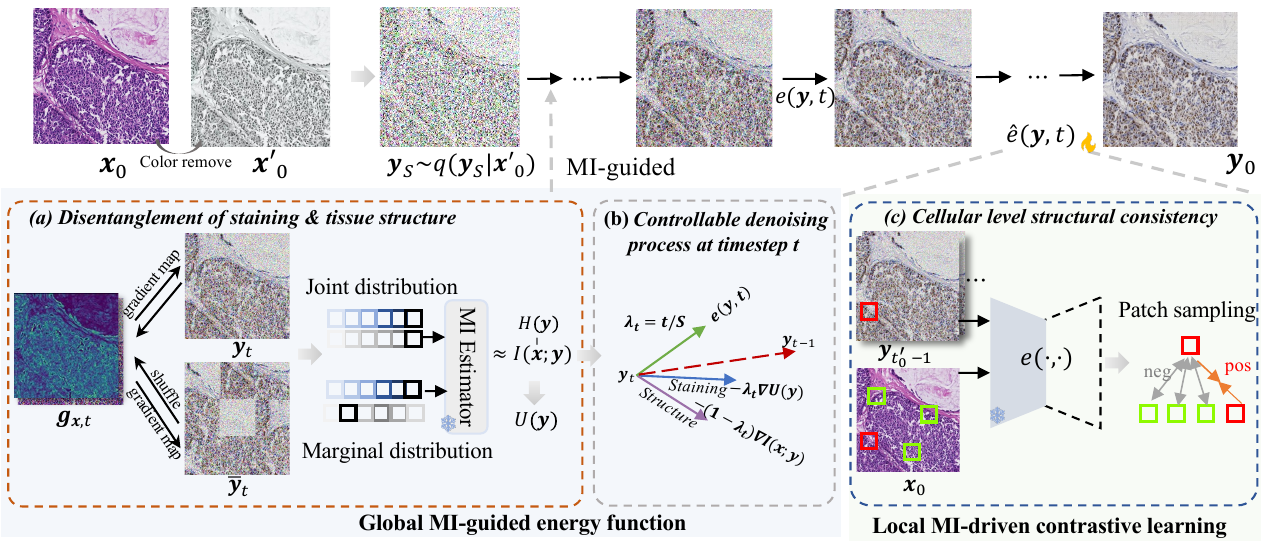}
    \caption{The overall workflow of MIU-Diff, including (a) disentanglement of the tissue structure and staining characteristics based on the MI-guided energy function, (b) controllable denoising process at each timestep with the proposed energy function, and (c) local MI-driven contrastive learning that ensures cross-modal structure consistency. 
    }
    \label{fig1}
\end{figure}

\section{Methodology}
Fig. \ref{fig1} illustrates the framework of MIU-Diff for unpaired histology staining. Given an H\&E-stained image $\textbf{\textit{x}}_0$, MIU-Diff aims to generate its IHC-stained counterpart $\textbf{\textit{y}}_0$. To address the unpaired nature of training data of H\&E and IHC images, MIU-Diff adopts a novel two-stage score-based diffusion model \cite{song2020score}, where it first learns the IHC marginal distribution $q(\textbf{\textit{y}}_{0})$ and then the conditional distribution $q(\textit{\textbf{y}}_{0}|\textit{\textbf{x}}_{0})$.
The main challenges in unpaired H\&E-to-IHC translation include disentangling staining style from tissue structure, ensuring controllability in the staining process, and maintaining cross-modal structural consistency. To tackle these, our framework (Section \ref{sec2.1}) incorporates a global MI-guided energy function (Section \ref{sec2.2}), a timestep-adaptive reverse-time SDE (Section \ref{sec2.2}), and a local MI-guided contrastive learning strategy (Section \ref{sec2.4}).

\subsection{Two-stage score-based diffusion framework}\label{sec2.1}
The proposed MIU-Diff enables unpaired virtual staining from source H\&E images to target IHC images through a two-stage diffusion process. In the first stage, the IHC distribution $q(\textbf{\textit{y}}_{0})$ is pretrained with an unconditional score-based diffusion model based on IHC images alone: The forward diffusion process $\{\textbf{\textit{y}}_{t}\}_{t\in[0,T]}$ follows a forward SDE \cite{song2020score}, where $\text{d}\textbf{\textit{y}} = -\frac{1}{2}\beta(t)\text{d}t+\sqrt{\beta(t)}\text{d}\textbf{\textit{w}}$, with $\beta(t), t\in [0,T]$ and $\textbf{\textit{w}}$  denoting the noise variance schedule and the standard Wiener process, respectively. Meanwhile, the reverse process is modeled using a variance-preserving (VP)-SDE: $\text{d}\textbf{\textit{y}}=[-\frac{1}{2}\beta(t)\text{d}t-\beta(t)e(\textbf{\textit{y}},t)]\text{d}t+\sqrt{\beta(t)}\text{d}\bar{\textbf{\textit{w}}}$, where $\bar{\textbf{\textit{w}}}$ is a standard Wiener process, $e(\textbf{\textit{y}},t)$ is a score-based model trained for approximating $\nabla_{y}\text{log}q_{t}(\textbf{\textit{y}})$, and $q_{t}(\textbf{\textit{y}})$ is the marginal distribution at timestep $t$.


Then, in the second stage, to effectively learn $q(\textbf{\textit{y}}_{0}|\textbf{\textit{x}}_{0})$ for unpaired virtual staining, we introduce an MI-guided energy function in the reverse process to disentangle staining style and tissue structure. Specifically, following \cite{wang2023stylediffusion}, we first remove the color from $\textbf{\textit{x}}_{0}$ to obtain $\textbf{\textit{x}}'_{0}$, as H\&E staining information is not required in virtual staining translation.
Given the start point $S$ from the noisy image $\textbf{\textit{y}}_{S}\sim q(\textbf{\textit{y}}_{s}|\textbf{\textit{x}}'_{0})$, the unpaired virtual staining process is guided as:
\begin{equation}\label{e1}
    \text{d}\textbf{\textit{y}}=[-\frac{1}{2}\beta(t)\text{d}t-\beta(t)(\hat{e}(\textbf{\textit{y}},t)-\nabla _{y}\mathcal{M}(\textbf{\textit{y}},\textbf{\textit{x}}_{0},t))]\text{d}t+\sqrt{\beta(t)}\text{d}\bar{\textbf{\textit{w}}},
\end{equation}
where $\mathcal{M}(\textbf{\textit{y}},\textbf{\textit{x}}_{0},t)$ is the MI-guided energy function that directs the reverse process to disentangle the structural and staining characteristics among H\&E images and generated IHC images.
Additionally, we solve the timestep-adaptive reverse-time SDE (Eq. \ref{e1}) for controllable virtual staining generation. Moreover, for local structural consistency across modalities, we incorporate a local MI-guided contrastive learning strategy. Specifically, $e(\textbf{\textit{y}},t)$ is optimized to obtain $\hat{e}(\textbf{\textit{y}},t)$ through patch-level contrastive learning during the later stages (after $t=t'_{0}$, $t'_{0}<\frac{1}{2}S$) of the reverse process. 

Details in the second stage are further introduced as follows.

\begin{figure}[t]
    \centering
    \includegraphics[width=0.85\linewidth]{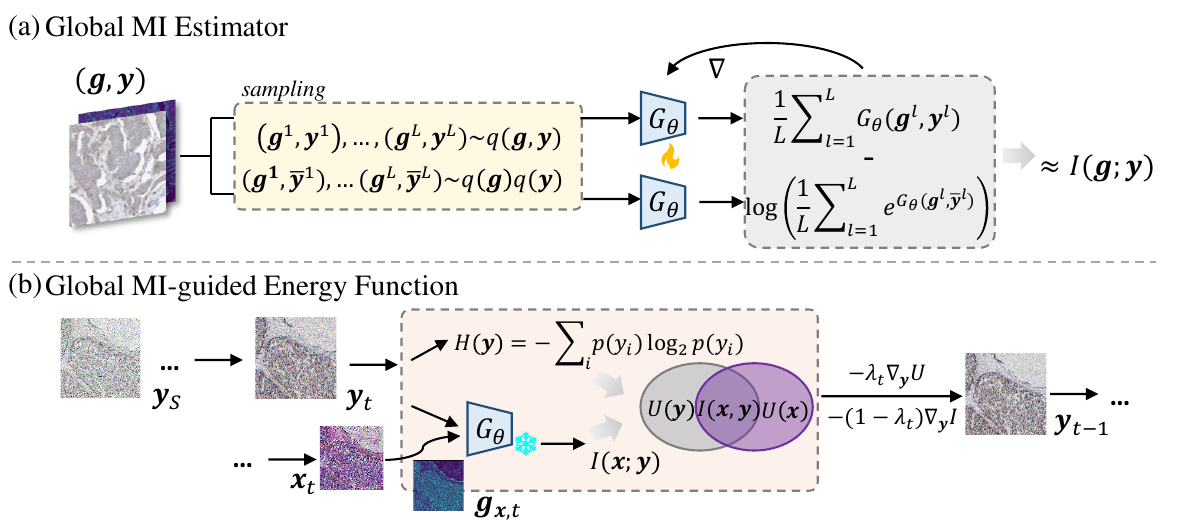}
    \caption{Illustration of (a) the construction and training process of the global MI estimator, and (b) the MI-guided energy function in the reverse diffusion, which is formulated by $I(\textbf{\textit{x}};\textbf{\textit{y}})$ approximated using the MI estimator and $U(\textbf{\textit{y}})$.}
    \label{fig2}
\end{figure}

\subsection{MI-guided energy function to disentangle staining \& structure}\label{sec2.2}
This section details the design of the MI estimator for calculating mutual information across modalities and its application in solving the timestep-adaptive reverse-time SDE. This specially designed reverse diffusion process dynamically guides IHC image generation, ensuring staining specificity while preserving global structural integrity from the source H\&E image.


\noindent\textbf{Global MI estimator.} As shown in Fig. \ref{fig2}(a), we design an MI estimator to quantify the shared information, i.e., structural characteristics, between the H\&E and its generated IHC image. Due to the misalignment of the existing H\&E-IHC pairs, we train the MI estimator $G_{\theta}$ within the IHC domain, i.e., learning to measure the shared information $I(\textbf{\textit{g}}_{\textbf{\textit{y}}},\textbf{\textit{y}})$ between the IHC image $\textbf{\textit{y}}$ and its gradient map\footnote{The gradient map can capture the structural information of an image \cite{ma2020structure}.} $\textbf{\textit{g}}_{\textbf{\textit{y}}}$.  Specifically, inspired by MINE \cite{belghazi2018mine}, given $L$ patch-level\footnote{Here, each IHC image and its gradient map are both cropped into $L$ patches.}  IHC-gradient pairs  $\left(\textbf{\textit{g}}_{\textbf{\textit{y}}}^{1},\textbf{\textit{y}}^{1}\right), \dots, \left(\textbf{\textit{g}}_{\textbf{\textit{y}}}^{L},\textbf{\textit{y}}^{L}\right) \sim q(\textbf{\textit{g}}_{\textbf{\textit{y}}},\textbf{\textit{y}})$ as joint distribution, and $L$ patch-level shuffled pairs $\left(\textbf{\textit{g}}_{\textbf{\textit{y}}}^{1},\bar{\textbf{\textit{y}}}^{1}\right), \dots, \left(\textbf{\textit{g}}_{\textbf{\textit{y}}}^{L},\bar{\textbf{\textit{y}}}^{L}\right) \sim q(\textbf{\textit{g}}_{\textbf{\textit{y}}})q(\textbf{\textit{y}})$ as marginal distribution, we train $G_{\theta}$ based on the principle that greater divergence between the joint distribution and the product of the marginal distributions indicates stronger dependence between $\textbf{\textit{y}}$ and $\textbf{\textit{g}}_{\textbf{\textit{y}}}$, with the specific. 
learning process formulated as  $I(\textbf{\textit{g}}_{\textbf{\textit{y}}};\textbf{\textit{y}})\ge I_{\theta }(\textbf{\textit{g}}_{\textbf{\textit{y}}};\textbf{\textit{y}})=\frac{1}{L}\sum_{l=1}^{L}G_{\theta }\left(\textbf{\textit{g}}_{\textbf{\textit{y}}}^{l},\textbf{\textit{y}}^{l}\right)-\log\left(\frac{1}{L}\sum_{l=1}^{L} e^{G_{\theta }\left(\textbf{\textit{g}}_{\textbf{\textit{y}}}^{l},\bar{\textbf{\textit{y}}}^{l}\right)} \right)$.

Then, in the reverse diffusion process, the MI $I(\textbf{\textit{x}};\textbf{\textit{y}}_t)$ between the generated target image $\textbf{\textit{y}}_t$ and source image $\textbf{\textit{x}}$ can be simply obtained using the well-trained MI estimator  $G_{\theta}$, which estimates the MI between  $\textbf{\textit{y}}_t$ and the gradient map of $\textbf{\textit{x}}_t$. Then, the final unique information of the staining information  in $\textbf{\textit{y}}_t$ can be ontained by $U(\textbf{\textit{y}}_t)=H(\textbf{\textit{y}}_t)-I(\textbf{\textit{x}}_t;\textbf{\textit{y}}_t)$, where $H(\textbf{\textit{y}}_t)$ is the entropy of $\textbf{\textit{y}}_t$.

\noindent\textbf{Solving timestep-adaptive reverse-time SDE for controllable generation.} 
As shown in Fig. \ref{fig2}(b), based on the $I(\textbf{\textit{x}};\textbf{\textit{y}})$ and $U(\textbf{\textit{y}})$ above, we design an MI-guided energy function to prioritize the restoration of specific staining details before shared structural information, inspired by \cite{choi2021ilvr}. Formally, the MI-guided energy function in the reverse-time SDE (Eq. \ref{e1}) is defined as
\begin{equation}\label{e2}
\begin{split}
    \mathcal{M}(\textbf{\textit{y}},\textbf{\textit{x}},t)=&\lambda_{t}\mathcal{U}(\textbf{\textit{y}},\textbf{\textit{x}},t)+(1-\lambda_{t} )\mathcal{I}(\textbf{\textit{y}},\textbf{\textit{x}},t)\\
    =&-\lambda_{t}\mathbb{E}_{q_{t|0}(\textbf{\textit{x}}_{t}|\textbf{\textit{x}})}U(\textbf{\textit{y}},\textbf{\textit{x}}_{t},t)-(1-\lambda _{t}
)\mathbb{E}_{q_{t|0}(\textbf{\textit{x}}_{t}|\textbf{\textit{x}})}I(\textbf{\textit{y}},\textbf{\textit{x}}_{t},t),
\end{split}
\end{equation}
where $\lambda_{t}=\frac{t}{S}$. Notably, the perturbation kernel $q_{t|0}(\cdot,\cdot)$ ensures the smoothness of the energy function, satisfying the global Lipschitz condition \cite{song2020score}.

Then, we further introduce the process in solving the reverse-time SDE (Eq. \ref{e1}) involving the MI-guided energy function (Eq. \ref{e2}). Inspired by \cite{song2020score}, we solve the reverse-time SDE based on the Euler-Maruyama method \cite{mao2015truncated}. Specifically, given a relatively small timestep of $\triangle = h-t$, the transition kernel of our reverse diffusion process $\hat{p}(\textbf{\textit{y}}_{t}|\textit{\textbf{y}}_{h})$ can be further represented using the Bayes' theorem as 
$\hat{p}(\textbf{\textit{y}}_{t}|\textit{\textbf{y}}_{h})\propto p(\textit{\textbf{y}}_{t}|\textit{\textbf{y}}_{h})p_{\mathcal{M}}(\textit{\textbf{y}}_{t}|\textit{\textbf{x}}_{0})$, where $p(\textbf{\textit{y}}_{t}|\textbf{\textit{y}}_{h})=\mathcal{N}(\boldsymbol{\mu},\boldsymbol{\Sigma} )$ represents
the transition kernel of the pre-trained diffusion model in the first stage, and  $p(\textit{\textbf{y}}_{t}|\textit{\textbf{y}}_{h})$ can be estimated based on  the MI-guided energy function. 
Assuming that $\mathcal{M}(\textbf{\textit{y}}_{t},\textbf{\textit{x}}_{0},t)$ has lower curvature compared to $\boldsymbol{\Sigma}^{-1}$, $p_{\mathcal{M}}(\textbf{\textit{y}}_{t}|\textbf{\textit{x}}_{0})$, it can be approximated by the Taylor expansion at $\textbf{\textit{y}}_{t}=\boldsymbol{\mu}$: Let $\textbf{\textit{k}}=\bigtriangledown _{\textbf{\textit{y}}'}\mathcal{M}(\textbf{\textit{y}}',\textbf{\textit{x}}_{0},t)|_{\textbf{\textit{y}}'=\boldsymbol{\mu}}$, we can get $\mathcal{M}(\textbf{\textit{y}}_{t},\textbf{\textit{x}}_{0},t)\approx \mathcal{M}(\boldsymbol{\mu},\textbf{\textit{x}}_{0},t)+(\textbf{\textit{y}}_{t}-\boldsymbol{\mu})^{\text{T}}\textbf{\textit{k}}$. Therefore, 
\begin{equation}\label{e3}
\begin{split}
    \log(\hat{p}(\textbf{\textit{y}}_{t}|\textbf{\textit{y}}_{h})) \approx &-\frac{1}{2}(\textbf{\textit{y}}_{t}-\boldsymbol{\mu}+\boldsymbol{\Sigma} \textbf{\textit{k}}) ^{\text{T}}\boldsymbol{\Sigma}^{-1}(\textbf{\textit{y}}_{t}-\boldsymbol{\mu}+\boldsymbol{\Sigma} \textbf{\textit{k}})+\text{Constant}\\
    =& \log(\textbf{\textit{b}})+\text{Constant}, \textbf{\textit{b}} \sim \mathcal{N}(\boldsymbol{\mu}-\boldsymbol{\Sigma} \textbf{\textit{k}}, \boldsymbol{\Sigma}),
\end{split}
\end{equation}
which means that the MI-guided reverse process can be achieved by shifting the mean of the transition kernel of the diffusion pre-trained in the IHC domain. Of note, the generation controllability lies in the flexibility in setting the total timesteps $N$ in which the MI-guided energy function is utilized.

\subsection{Local MI-driven contrastive learning}\label{sec2.4}
To further ensure structural consistency between the generated IHC and source H\&E images, inspired by \cite{park2020contrastive}, we develop a local MI-driven contrastive learning strategy to optimize the diffusion model (illustrated in Fig. \ref{fig1}(c)).
Specifically, in the reverse diffusion process, starting from the initial timestep $S$ to the intermediate timestep $t'_{0}$
\footnote{the settings of intermediate timestep $t'_{0}$ are explored in Table \ref{tab2}.}, we first apply reverse-time SDE using the pretrained score-based model to roughly preserve the tissue structure of $\textbf{\textit{x}}_{0}$. Next, we further refine the local structural details of the generated target image $\textbf{\textit{y}}_{0}$.

From $t'_{0}$ onward, we update the score-based model $e(\textbf{\textit{y}},t)$ using a patch-wise contrastive loss defined as $\ell_{PCL}(\textbf{\textit{y}}_{t},\textbf{\textit{x}}_{0})= \mathbb{E}_{\textbf{\textit{y}}_{t},t}\left[\sum_{i}\sum_{j} \ell(\textbf{\textit{z}}_{\textbf{\textit{y}}_{t},i}^{p},\textbf{\textit{z}}_{\textbf{\textit{x}}_{0},i}^{p},\textbf{\textit{z}}_{\textbf{\textit{x}}_{0},i}^{P\setminus p}) \right]$, where $\ell$ is the cross-entropy loss, and $\textbf{\textit{z}}_{\textbf{\textit{y}}_{t},i}^{p}$ and $\textbf{\textit{z}}_{\textbf{\textit{x}}_{0},i}^{p}$ are the encoder features from the $i$-th layer at patch position $p$ of $\textbf{\textit{y}}_{t}$ and $\textbf{\textit{x}}_{0}$, respectively. Patches are randomly sampled from the location set  $P$. Corresponding patches at position $p$  in $\textbf{\textit{z}}_{\textbf{\textit{y}},i}$ and $\textbf{\textit{z}}_{\textbf{\textit{x}}_0,i}$ form "positive" pairs, while non-corresponding patches $\textbf{\textit{z}}_{\textbf{\textit{x}},i}^{P\setminus p}$ form "negative" pairs. The objective of $\ell_{PCL}$ is to maximize mutual information between positive patches while minimizing it for negative patches.

\section{Experiments}
\noindent\textbf{Datasets.} We evaluate our method on two public datasets: the Breast Cancer Immunohistochemical (BCI) challenge dataset \cite{liu2022bci} and the Multi-IHC Stain Translation (MIST) dataset \cite{vsgan3}. In BCI, we split the dataset into 3498, 437 and 438  H\&E-IHC (of gene HER2) pairs for training, validation and test. For the MIST dataset, we include  H\&E-IHC pairs of two genes, i.e., HER2 (4,513, 564 and 563 pairs for training, validation and test) and Ki67 (4,288, 536, 537 pairs for training, validation and test). All patches
are cropped in size of 1024$\times$1024. All H\&E-IHC pairs are from consecutive slices, but not strictly pixel-aligned.

\noindent\textbf{Implementation details.} Our model is implemented  using Pytorch on two Tesla V100 32GB GPUs. The pretrained diffusion model in the IHC domain is trained for 40k iterations with batch size 4 and learning rate $1\times10^{-4}$ using AdamW optimizer. The images are resized from 1024$\times$1024 to 256$\times$256 due to memory limitations. For validation and test, the batch size was 1, and the hyper-parameter $N$ and $t'_{0}$ are set to 300 and 40, respectively.

\subsection{Performance evaluation}
\begin{table}[t]
\centering
\caption{Performance of virtual IHC staining. Bold numbers indicate the best results.}
\resizebox{0.98\textwidth}{!}{
\begin{tabular}{cccccccccccccccc}
\hline
\multirow{2}{*}{Method} &  & \multicolumn{4}{c}{$\text{MIST}_{\text{HER2}}$}                                      &  & \multicolumn{4}{c}{$\text{MIST}_{\text{Ki67}}$}                                      &  & \multicolumn{4}{c}{$\text{BCI}_{\text{HER2}}$}                                       \\ \cline{3-6} \cline{8-11} \cline{13-16} 
                        &  & PSNR$\uparrow$            & VIF$\uparrow$            & Hist$\uparrow$           & IOD$\times 10^{7}$         &  & PSNR$\uparrow$            & VIF$\uparrow$            & Hist$\uparrow$           & IOD$\times 10^{7}$         &  & PSNR$\uparrow$            & VIF$\uparrow$            & Hist $\uparrow$          & IOD$\times 10^{7}$        \\ \hline
CUT\cite{park2020contrastive}                     &  & 13.123          & 0.610          & 0.264          & -1.168          &  & 13.103          & \textbf{0.611} & 0.331          & -2.705          &  & 17.649          & 0.762          & 0.571          & -0.825          \\
ASP \cite{li2023adaptive}                     &  & \textbf{15.069} & 0.601          & 0.495          & -1.325          &  & \textbf{14.339} & 0.609          & 0.429          & -1.604          &  & \textbf{18.336} & 0.732          & 0.513          & -0.804          \\ \hline
ILVR\cite{choi2021ilvr}                     &  & 12.231          & 0.598          & 0.147          & -3.456          &  & 13.501          & 0.509          & 0.142          & -2.650          &  & 13.459          & 0.497          & 0.299          & -2.637          \\
SDEdit\cite{meng2021sdedit}                   &  & 12.054          & 0.599          & 0.202          & -2.463          &  & 13.760          & 0.529          & 0.270          & -1.611          &  & 12.393          & 0.513          & 0.493          & -1.280          \\
EGSDE\cite{zhao2022egsde}                    &  & 13.607          & 0.587          & 0.225          & -2.548          &  & 12.447          & 0.468          & 0.361          & -3.637          &  & 12.836          & 0.603          & 0.379          & -1.837          \\ \hline
\textit{\textbf{Ours}}  &  & 14.368          & \textbf{0.611} & \textbf{0.670} & \textbf{-1.146} &  & 14.201          & 0.586          & \textbf{0.507} & \textbf{-0.523} &  & 15.120          & \textbf{0.773} & \textbf{0.640} & \textbf{-0.792} \\ \hline
$I$-guided           &  & 14.293          & 0.609          & 0.321          & -1.903          &  & 13.503          & 0.482          & 0.334          & -2.022          &  & 15.068          & 0.713          & 0.596          & -1.631          \\
$U$-guided           &  & 13.878          & 0.600          & 0.558          & -2.095          &  & 13.718          & 0.570          & 0.491          & -1.732          &  & 14.597          & 0.685          & 0.630          & -2.215          \\
$w/o$ MI-guided           &  & 14.337          & 0.589          & 0.217          & -2.178           &  & 13.055          & 0.480          & 0.320          & -2.998          &  & 15.062          & 0.659          & 0.561          & -2.887          \\ \hline
$\ell_{2}$ optimization         &  & 13.226          & 0.591          & 0.403          & -1.867          &  & 13.433          & 0.553          & 0.440          & -2.670          &  & 13.809          & 0.698          & 0.614          & -2.825          \\
$w/o$ $\ell_{PCL}$                 &  & 12.932          & 0.588          & 0.663          & -2.311          &  & 12.989          & 0.525          & 0.505          & -2.938          &  & 14.093          & 0.643          & 0.621          & -2.451          \\ \hline
\end{tabular}}\label{tab1}
\end{table}

\begin{figure}[t!]
    \centering
    \includegraphics[width=0.82\linewidth]{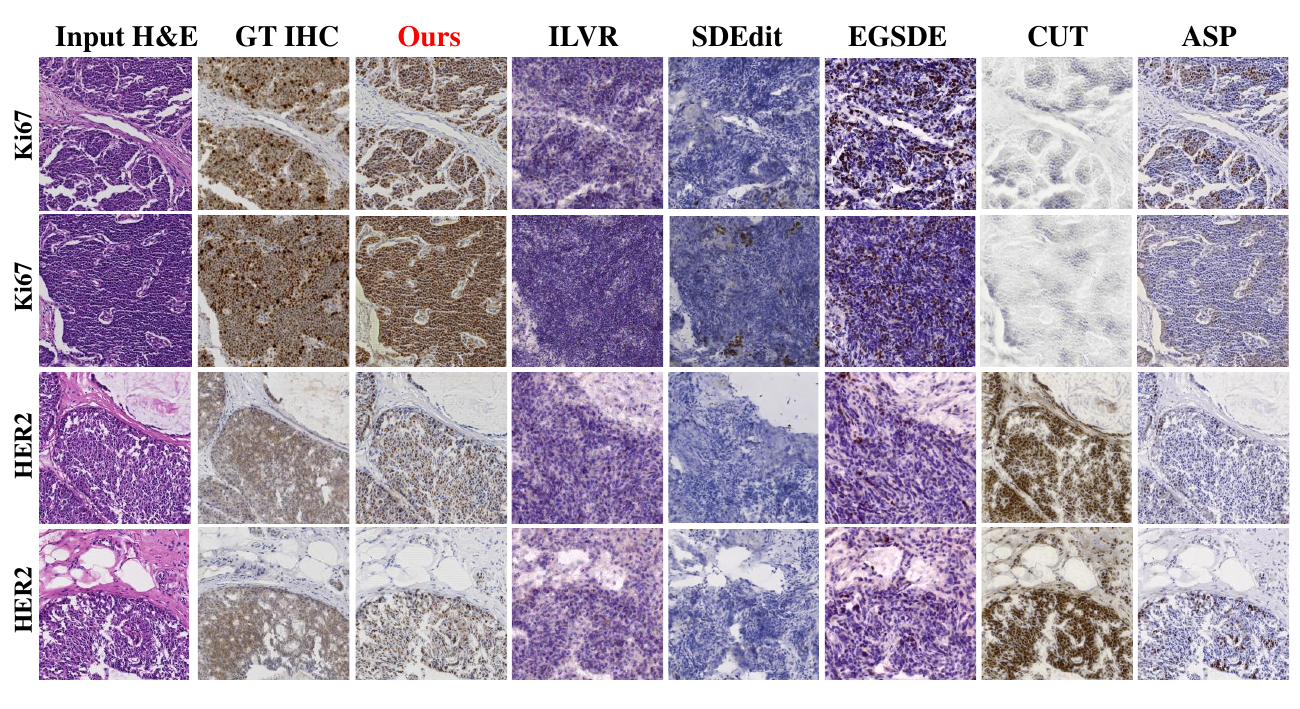}
    \caption{Visual comparisons on $\text{MIST}_{\text{Ki67}}$ and $\text{MIST}_{\text{HER2}}$.}
    \label{fig3}
\end{figure}
We compare our model with advanced image translation methods, including three diffusion-based models: ILVR \cite{choi2021ilvr}, SDEdit \cite{meng2021sdedit}, EGSDE \cite{zhao2022egsde} and two GAN-based methods: CUT \cite{park2020contrastive} and ASP \cite{li2023adaptive}. To assess the quality of the generated IHC images, we use the objective assessment metric peak signal-to-noise ratio (PSNR) and subjective metrics: visual information fidelity (VIF) for assessing information preservation \cite{sheikh2006image}, and histogram correlation (Hist) for evaluating color matching performance. Moreover, to evaluate the IHC virtual staining performance in clinically relevant areas, following \cite{zhang2022mvfstain}, we calculate the integrated optical density (IOD) deviation between the generated IHC and its ground truth (GT) in segmented cellular regions. Experimental results are tabulated in Table \ref{tab1} with 300 denoising timesteps for all diffusion-based models for fairness: MIU-Diff achieves the best performance in information fidelity and color matching. Furthermore, MIU-Diff shows the smallest deviation in IOD from GT, indicating the highest consistency in positive signals at the cellular level. Notably, a higher PSNR in unpaired virtual staining tasks does not always indicate better generation quality, as mismatched structures between H\&E and IHC images prevent pixel alignment. Fig. \ref{fig3} provides visual comparisons on $\text{MIST}_{\text{Ki67}}$ and $\text{MIST}_{\text{HER2}}$, where MIU-Diff outperforms all other methods, maintaining structural consistency while achieving accurate staining transformation.


\subsection{Ablation Study and hyper-parameter sensitivity analysis}
Here, we assess the contribution of key components in MIU-Diff. Specifically, for global MI-guided energy function, we evaluate: 1) $I$-guided - using only $I(\textbf{\textit{x}};\textbf{\textit{y}})$ guidance; 2) $U$-guided - using only $U(\textbf{\textit{y}})$ guidance; 3) $w/o$ MI-guided - without any guidance in the reverse process. As shown in Table \ref{tab1}, without guidance for structure and staining leads to loss of structural information and color distortion. For the local MI-driven contrastive learning, replacing it with $\ell_2$ loss results in the loss of structural details and reduction in staining quality; Removing $\ell_{PCL}$ ($w/o$ $\ell_{PCL}$) leads to a decrease in positive signal consistency with GT. Table \ref{tab2} further explores the impact of $t'_{0}$ and  $N$, which decide the timesteps for the $\ell_{PCL}$ and the MI-guided energy function, respectively: Reducing $N$ facilitates retaining structural details, but excessively small $N$ (e.g., $N=100$) deteriorates staining restoration, causing distortion and impaired positive signal recognition; Larger $t'_{0}$ improves local structural consistency but increase computational cost. Moreover, $t'_{0} = 80$ causes image sharpening and staining bias.

\begin{table}[]
\caption{Ablation experiments with different optimization and denoising steps.}
\centering
\resizebox{0.7\textwidth}{!}{
\begin{tabular}{cccccccc}
\noalign{\hrule height 0.3mm} \vspace{-0.2cm}
                                                  &      &                &                 &                             &        &                 &                                     \\   
\multicolumn{1}{c|}{}                          &      & $t'_{0}$=80           & $t'_{0}$=40 (\textbf{\textit{Ours}})            & \multicolumn{1}{c|}{$t'_{0}$=10}   & $T$=500  & $T$=300 (\textbf{\textit{Ours}})           & \multicolumn{1}{c}{$T$=100}          \\ \cline{2-8} 
\multicolumn{1}{c|}{\multirow{3}{*}{$\text{MIST}_{\text{HER2}}$}} & VIF$\uparrow$  & \textbf{0.613} & 0.611           & \multicolumn{1}{c|}{0.597}  & 0.581  & 0.611           & \multicolumn{1}{c}{\textbf{0.616}} \\
\multicolumn{1}{c|}{}                            & Hist$\uparrow$ & 0.548          & \textbf{0.670}  & \multicolumn{1}{c|}{0.667}  & 0.584  & \textbf{0.670}  & \multicolumn{1}{c}{0.480}          \\
\multicolumn{1}{c|}{}                            & IOD$\times 10^{7}$  & -1.280         & \textbf{-1.146} & \multicolumn{1}{c|}{-1.773} & -1.993 & \textbf{-1.146} & \multicolumn{1}{c}{-2.037}         \\ \noalign{\hrule height 0.3mm} \vspace{-0.2cm}
                                                  &      &                &                 &                             &        &                 &                                     \\
\multicolumn{1}{c|}{\multirow{3}{*}{$\text{MIST}_{\text{Ki67}}$}} & VIF$\uparrow$  & \textbf{0.588} & 0.586           & \multicolumn{1}{c|}{0.561}  & 0.563  & \textbf{0.586}  & \multicolumn{1}{c}{0.585}          \\
\multicolumn{1}{c|}{}                            & Hist$\uparrow$ & 0.459          & \textbf{0.507}  & \multicolumn{1}{c|}{0.483}  & 0.474  & \textbf{0.507}  & \multicolumn{1}{c}{0.341}          \\
\multicolumn{1}{c|}{}                            & IOD$\times 10^{7}$  & -1.980         & \textbf{-0.523} & \multicolumn{1}{c|}{-2.330} & -2.937 & \textbf{-0.523} & \multicolumn{1}{c}{-2.441}         \\ \noalign{\hrule height 0.3mm} \vspace{-0.2cm}
                                                  &      &                &                 &                             &        &                 &                                     \\
\multicolumn{1}{c|}{\multirow{3}{*}{$\text{BCI}_{\text{HER2}}$}}  & VIF$\uparrow$  & \textbf{0.787} & 0.773           & \multicolumn{1}{c|}{0.698}  & 0.727  & 0.773           & \multicolumn{1}{c}{\textbf{0.784}} \\
\multicolumn{1}{c|}{}                            & Hist$\uparrow$ & 0.590          & \textbf{0.640}  & \multicolumn{1}{c|}{0.630}  & 0.596  & \textbf{0.640}  & \multicolumn{1}{c}{0.224}          \\
\multicolumn{1}{c|}{}                            & IOD$\times 10^{7}$  & -0.961         & \textbf{-0.792} & \multicolumn{1}{c|}{-2.107} & -2.653 & \textbf{-0.792} & \multicolumn{1}{c}{-2.218}         \\ \noalign{\hrule height 0.3mm}
\end{tabular}}\label{tab2}
\end{table}

\section{Conclusion}

The H\&E-IHC virtual staining task faces challenges in structural consistency preservation and staining conversion. In this study, we proposed an MI-guided score-based diffusion model, MIU-Diff, for unpaired virtual staining. MIU-Diff effectively disentangles staining style from tissue structure through a dynamic global MI-guided energy function. Additionally, the timestep-customized reverse diffusion process allows precise modulation of staining and structure. To further ensure cellular-level structural consistency, we also introduced a local MI-driven contrastive learning strategy. Experimental results on two public datasets showed superior performance of MIU-Diff over advanced methods, making it a valuable technique for clinical diagnosis.




%
%
%
 \clearpage
\bibliographystyle{splncs04}
\bibliography{mybibliography}

\end{document}